\documentstyle[12pt]{article}
\newcommand{\be}{\begin{equation}}
\newcommand{\ee}{\end{equation}}
\newcommand{\s}{\section}
\newcommand{\ci}{\cite}

\begin{document}
\begin{titlepage}

\begin{center}

\vspace{1.0cm}
\vskip 1.0cm
{\Large {\bf Nucleon and delta masses in $QCD$ $^*$}}\\

\vskip 2.5cm

{\large M. Rafecas $^{**}$ and  V.Vento $^{\ddag}$}

\vskip 0.2cm
{Departament de F\'{\i}sica Te\`{o}rica and I.F.I.C.}\\
{Centre Mixt Universitat de Val\`{e}ncia -- C.S.I.C.}\\
{E-46100 Burjassot (Val\`{e}ncia), Spain.}

\vspace{2.0cm}
{\bf Abstract}
\vspace{0.1cm}

\begin{quotation}
{\small Using the positivity of the path integral measure of
$QCD$ and defining a structure for the quark propagator in a
background field according to the fluxon scenario for confinement,
we calculate and compare the correlators for nucleon and delta. From
their shape we elucidate about the origin of their mass
difference, which in our simplified scenario is due to
the tensor  structure in the propagator. This term
arises due to a dynamical mechanism which is responsible
simultaneously for confinement and spontaneous chiral symmetry breaking.
Finally we discuss, by comparing the calculated correlators with the
Lehmann representation, the possibility that a strong CP and/or P violation
occurs as a consequence of a specific mechanism for confinement.}

\end{quotation}

\end{center}
\vspace{4.0cm}
$^*$ Supported in part by CICYT grant \# AEN90-0040 and DGICYT
grant \# PB88-0064. \\
$^{**}$ Doctoral Fellow of CSIC-Formaci\'on Profesorado Universitario \\
$^{\ddag}$ Rafecas@evalvx; Vento@evalun11; Vento@vm.ci.uv.es
\end{titlepage}
\baselineskip  0.2in

\s{Introduction}
It was realized long after $QCD$ was formulated that one could
derive some exact inequalities between hadron masses
\ci{inequalities} and other observables \ci{inequalities1}.
The key element in deriving them is that the
Euclidean fermion determinant in vector like gauge theories (such
as $QCD$) is positive definite and so the measure
\be
d\mu = Z^{-1} D A^a_{\mu} (x) det (i{\not\!\! D} + M)
\exp{(-\frac{1}{2g^2} \int d^4x TrF_{\mu \nu}^2)}
\ee
for the $A^a_{\mu}$ integration obtained after integrating out
the fermions is positive definite for $\Theta = 0$. Note that
${\not\!\! D} = \gamma^{\mu} D_{\mu}$, $D_{\mu}$ being the covariant
derivative. Inequalities that hold pointwise continue to hold
after integrating with respect to a positive measure. Thus any
inequality among matrix elements that holds after performing the
Fermi integral in a fixed background gauge field holds in the
exact theory.

Mass inequalities have been obtained among the mesons and
comparing baryons with mesons \ci{inequalities}. The important
property in these calculations has been
\be
S^+(x,y) =\gamma_5 S(y,x) \gamma_5
\ee
where $S(x,y)$ is the quark propagator in a background. Our aim
is to discuss mass relations among baryons, but due to their
current quark constituency the previous property is of no use.
Some fine details of the structure of the quark propagator and
of the baryon currents will be necessary to be able to address
the issue.

We center our investigation in the nucleon--delta mass difference.
This observable has become the first test of any hadron model, and
they all agree by now that it is different from zero, even if the
desintegration channel for the delta is decoupled. Therefore both
particles shall be treated on an equal footing in our work.

The procedure to follow is straight forward. We calculate the
baryon correlators for these two systems and study their
asymptotic behavior which is dominated exponentially by their
masses \ci{inequalities}. To do so we describe the baryon fields
in terms of elementary quark fields \`a la Ioffe \ci{ioffe}.
Some plausible assumptions on the structure of the quark
propagator are required in order to be predictive. We will use
in this paper the continuum language of Witten \ci{inequalities}.
To preserve the positivity of the fermion determinant for a fixed
$A^a_{\mu}$ and do the integral a suitable cut-off is required.
Recent developments give us confidence in the result of our
discussions \ci{asorey}.

\s{The baryon correlators}
The first step is to find composite gauge invariant operators
which have all the quantum numbers of the nucleon and the delta.
We choose the proton to perform the calculation. There are two
such  operators for it \ci{ioffe,espriu},
\be
O_1^{udu}(x) = \varepsilon^{abc}[u^T_a (x) {\it
C} d_b (x)]\gamma_5 u_c (x)
\ee
and

\be
O_2^{udu}(x) = \varepsilon^{abc}[u^T_a (x) {\it
C}\gamma_5 d_b (x)] u_c (x)
\ee
Here $T$ means transposed and ${\it C}$ is the charge
conjugation operation. We take initially the general
combination
\be
O(x) = O_1(x) +t O_2(x)
\ee
where $t$ can be considered the tangent of a mixing angle.
However in our final discussion, we shall simplify our expressions
to  the chiral limit ($t = -1$), because it is favored by the data
\ci{espriu}.

For the delta states there is only one such operator. We work for
simplicity, with that corresponding to the $\Delta^{++}$, i.e.,
\be
O^{\mu}_{uuu} = -i \varepsilon^{abc}[u^T_a{\it
C}\gamma^{\mu}u_b]u_c
\ee
Once we substitute in the expression for the baryon fields the
correlators become
\[
<N_{\alpha}(x) \bar{N}_{\beta}(y)> \sim \int d\mu [(\gamma_5
S(x,y) \gamma_5)_{\alpha \beta}\; Tr(S(x,y) S(x,y)) \]
\[
\;\;\;\;\;\;\;\;\;\;\;\;\;\;\;\;\;\;\;\;\; +\;
t\; \{ S(x,y), \gamma_5\}_{\alpha \beta} \; Tr(S(x,y) S(x,y) \gamma_5)\]
\be
\;\;\;\;\;\;\;\;\;\;\;\;\;\;\;\;\;\;\;\;\; +\;
t^2\; S(x,y)_{\alpha \beta} \;Tr(S(x,y) \gamma_5 S(x,y) \gamma_5)]
\label{nucleonc}
\ee
and
\[
<\Delta^{\mu}_{\alpha}(x) {\bar \Delta}^{\nu}_{\beta}(y)> \sim \int
d\mu [ S_{\alpha \beta} Tr(\gamma^{\mu} S(x,y) \gamma^{\nu} S(x,y))
\]
\be
\;\;\;\;\;\;\;\;\;\;\;\;\;\;\;\;\;\;\;\;\;\;\; +\;
(S(x,y) \gamma^{\nu} S(x,y) \gamma^{\mu} S(x,y))_{\alpha \beta}]
\label{deltac}
\ee
where $\{\;,\;\}$ represents an anticommutator.
At large separation, where the spectrum becomes explicit, and
taking into account the observed symmetries of the strong
interactions, the structure of the correlator should be dominated
by the Lehmann decomposition \ci{lehmann,schweber}
\be
<N_{\alpha}(x) \bar{N}_{\beta}(y)> \;\; \longrightarrow \;\;
F_{\mu 1}(x-y) \gamma^{\mu}_{\alpha \beta} + F_2(x-y)
\delta_{\alpha \beta}
\label{lehmannn}
\ee
and
\be
<\Delta^{\mu}_{\alpha}(x) {\bar \Delta}^{\nu}_{\beta}(y)>
\;\; \longrightarrow \;\; g^{\mu \nu} ( F^{\Delta}_{1 \sigma}(x-y)
\gamma^{\sigma}_{\alpha \beta} + F^{\Delta}_2 (x-y) \delta_{\alpha
\beta}) + \ldots
\label{lehmannd}
\ee
The $F$ functions, which can be extracted from
Eqs. (\ref{nucleonc}) and (\ref{deltac}), by comparison with
Eqs.(\ref{lehmannn}) and (\ref{lehmannd}), contain information
about the exponential fall off, i.e., {\it pole} in momentum space,
and therefore on the masses. Among the tensorial structures which appear
in Eq.(\ref{lehmannd}) we are interested in the one proportional to
$g_{\mu \nu}$, because it is the only one in which the delta gives a
contribution to the {\it pole} but not the nucleon \ci{espriu}.

\s{The mass relations}
In the present approach the properties of the quark propagator
are crucial, not only to determine relations among various
observables, but also to establish the possible realization of
the different symmetries of the theory \ci{vafa,nussinov}.
The most general quark propagator in a background field has the
following structure
\be
S(x,y) = s(x,y) + p(x,y) \gamma_5 + v^{\mu}(x,y) \gamma_{\mu} +
a^{\mu}(x,y) \gamma_{\mu} \gamma_5 + t^{\mu \nu}(x,y) \sigma_{\mu \nu}
\ee
Configurations to the $p$, $a^{\mu}$ and $t^{\mu \nu}$ terms arise only
through complex field configurations: instantons, fluxons, etc...
\ci{nussinov,schwinger} and, as we shall see, their behavior is
restricted if $QCD$ is to reproduce the behavior described by the
Lehmann representation \ci{lehmann}.

The fluxon mechanism has been
proposed as a possible scenario where chiral symmetry breaking
could occur in a confining theory \ci{fluxons}. Recent Lattice
Montecarlo calculations point towards a mechanism similar to this
scenario \ci{pisa}. We take this ansatz, however, because it contains the
minimal structure necessary to present our ideas and cannot be naively
discarded. Under this circumstances the quark propagator
could have the following form
\ci{nussinov,schwinger} \footnote{We shall analyze the
effects of a pseudoscalar term, its relation with the anomaly
and some experimental consequences, in a future publication
\ci{rafecas}.}
\be
S(x,y) = s(x,y)  + v^{\mu}(x,y) \gamma_{\mu} +
t^{\mu \nu}(x,y) \sigma_{\mu \nu}
\ee
Calculating the correlation functions we obtain for the proton

\[<N{\bar N}> \sim \frac{1}{2}\int d\mu \{[(s^2 + t^{\alpha
\beta}t_{\alpha \beta}) (1+t^2) + v^2 (1-t^2)] s \]
\[\;\;\;\;\;\;\;\;\;\;\;\;+\; [(s^2 + t^{\alpha \beta}t_{\alpha \beta})
(1-t^2) + v^2 (1+t^2)] v^{\mu} \gamma_{\mu} \]
\be
\;\;\;\;\;\;\;\;\;\;\;\;\;\;\;+\; [(s^2 + t^{\alpha \beta}t_{\alpha \beta})
(1+t^2) + v^2 (1-t^2)] t^{\delta \epsilon} \sigma_{\delta \epsilon}\}
\label{nucleon}
\ee
and for the delta

\[ <\Delta^{\mu} \Delta_{\mu}> \sim \int d\mu \{s(\frac{4}{3}s^2-v^2 -
t^{\alpha \beta} t_{\alpha \beta}) - \frac{i}{2}
s\varepsilon_{\alpha \beta \gamma \delta} t^{\alpha \beta}
t^{\gamma \delta} \gamma_5\]
\[\;\;\;\;\;\;\;\;\;\;+\; [(\frac{5}{4}s^2-\frac{3}{4}v^2 + \frac{1}{2}
t^{\alpha \beta} t_{\alpha \beta})v^{\gamma} +
\frac{1}{2}t^{\alpha \beta} v_{\beta}
t^{\gamma}_{\alpha}]\gamma_{\gamma}\]
\[\;+\; [(s^2 - \frac{1}{2}v^2)t^{\alpha \gamma} - 2ist^{\alpha
\beta}t^{\gamma}_{\beta}] \sigma_{\alpha \gamma}\]
\be
\;\;\;\;\;\;\;\;\;\;\;-\;\frac{1}{2} \varepsilon_{\alpha \beta \gamma
\delta}[g^{\mu \delta}st^{\alpha \beta} v^{\gamma} + i
t^{\alpha \beta} t^{\delta \mu} v^{\gamma} + i t^{\alpha
\rho}v_{\rho} t^{\beta \gamma} g^{\mu \delta}]\gamma_{\mu}
\gamma_5 \}
\label{delta}
\ee
Recalling at this point the Lehmann representation it is apparent that
the above equations contain exotic terms not present in the
latter and which imply violations, hopefully small, of some of the
symmetries of the strong interactions \ci{schweber}.

It has been shown \ci{ioffe,espriu} that $t = -1$, the value in
the chiral limit is close to that preferred by the data. Let us study therefore
this limit and impose consistently the zero quark mass limit. The
above expressions simplify notably and we thus obtain
\be
F_2 = F^{\Delta}_2 = 0
\ee
\be
F^{\mu}_1 \sim  - \int d\mu v^2v^{\mu}
\ee
\be
F^{\Delta \mu}_1 \sim - \int d\mu (v^2 v^{\mu} + \frac{2}{3} v^{\mu}
t^{\alpha \beta} t_{\alpha \beta} + \frac{2}{3}t^{\alpha
\beta}v_{\beta}t^{\mu}_{\alpha})
\ee
As has been shown in ref.(\ci{nussinov}) a non vanishing tensor
component leads to the spontaneous breaking of chiral symmetry.
Thus the same mechanism which produces in our scenario chiral
symmetry breaking leads to the mass difference between the nucleon
and the delta \footnote{The pseudoscalar and scalar contributions
vanish in the massless limit if one assumes, as we do here, that no
nonperturbative phenomena associated with confinement occur in these
terms (see also discussion in \ci{nussinov}).
Moreover we work under the assumption that the tensor component
corresponds to the scenario of strong confining fields and does
not vanish \ci{nussinov,pisa}. Therefore spontaneous chiral symmetry
breaking is entirely described by the tensor component in this limit.}.

In the fluxon scenario, the pion and $\sigma$ correlators are given
by \ci{inequalities}
\be
<\pi \pi> \sim \int d\mu (|v|^2 + t^{\alpha \beta}t_{\alpha
\beta}^*)
\ee
\be
<\sigma \sigma> \sim \int d\mu (|v|^2 - t^{\alpha \beta}t_{\alpha
\beta}^*)
\ee
In the Wigner realization of chiral symmetry, only the vector
term does not vanish and the sigma and pion have the same mass.
In the Goldstone realization of chiral symmetry the tensor
component must cancel the exponential asymptotic behavior leading
to a constant one, i.e., a Goldstone pion. The tensor interaction,
which turns out to be attractive in the pion channel must be repulsive
in the delta channel. Comparing the Lehmann representation with
Eq.(\ref{delta}), we see that tensor terms must produce an
asymptotically subleading behavior and therefore
\be
\int d\mu |(v^2 v^{\mu} + \frac{2}{3} v^{\mu}
t^{\alpha \beta} t_{\alpha \beta} + \frac{2}{3}t^{\alpha
\beta} v_{\beta} t^{\mu}_{\alpha})| < {\it k} | \int d\mu
v^2v^{\mu}|
\ee
where {\it k} is a constant.
Thus the mass difference between the nucleon and the delta is
expressing the fact, that the tensor components, which produce
the leading behavior in the pionic sector, contribute destructively
in the baryonic sector generating a subleading behavior.
\s{Exotic terms}
Let us look at Eqs.(\ref{nucleon}) and (\ref{delta}) as obtained
from our calculation. As we have previously mentioned we
obtain heterodox components in our correlator, assuming
orthodoxy given by the Lehmann representation. In the chiral
limit for example
\be
Nucleon \sim \int d\mu [t^{\alpha \beta} t_{\alpha \beta}
t^{\delta \epsilon} \sigma_{\delta \epsilon}]
\ee
\be
Delta \sim \frac{1}{2 }\int d\mu [v^2 t^{\alpha \gamma} \sigma_{\alpha \gamma}
+ i\epsilon_{\alpha \beta \gamma \delta} (t^{\alpha \beta}
t^{\delta \mu}v^{\gamma} + t^{\alpha \rho}v_{\rho}t^{\beta \gamma}
g^{\mu \delta})
\gamma_{\mu} \gamma_5]
\ee
The deviations from the Lehmann representation due to a tensor structure
imply a strong violation of CP, while those of an axial vector structure
that of P \ci{schweber}.
We see that the exotic terms in the delta and in the nucleon do not correspond
to the same Lorentz structure, having the delta an additional axial vector
term. Although naive power counting of tensor and vector components might
indicate that the axial vector term is of the same order as
the non exotic vector contribution,the antisymmetric product is making
this non-observed contribution vanishingly small, thus giving a hint
about the structure of the fluxon mechanism. It is important to notice that
such a term might have experimental consequences and one should analyze
its implications. Besides it, both nucleon and delta contain
an exotic tensor structure. By looking at the non exotic vector components
we might infer that the magnitude in both cases must be similar, since the
tensor and vector parts have similar leading behaviors. Thus one can use
either nucleon or delta observables to detect this type of deviations
from the experimentally
dominant symmetry structures\footnote{ We have analyzed the
contribution from the pseudoscalar also \ci{rafecas} and an
analogous conclusion can be drawn. Concrete experimental
proposals for both cases are being investigated.}.

\s{Conclusion}

The fact that the $QCD$ measure is positive has been used to
establish exact relations among hadronic observables. These
results, although exact, are not  quantitative,
like those obtained by lattice calculations. Moreover, in
the case of baryons, due to their quark constituency, no relations
among observables has been established. In this work we initiate a new
line of thought, based on these ideas, but which requires from
some additional input in the form of the quark propagator. One
looses exactitude, but in principle one might get
quantitative predictions. We have applied the technique to the
baryon case, and by formulating a precise structure of the quark
propagator we have analyzed some of the implications of the
baryon spectrum into the behavior of the theory. We have not
aimed at quantitative results in this first attempt. One could
do so by postulating a precise model for the fermionic
propagator in terms by weak field expansions
\ci{schwinger}, semiclassical fields as hinted by lattice
calculations \ci{pisa} or postulating some stromg field limit,
and in so doing define models of hadron structure,
at the level of the full quantum theory. Here
we have only described immediate implications of the tensor structure
in the fermion propagator, which we next recall.

We have assumed a quark propagator as would
appear from a fluxon picture of confinement. We have taken a simplified
scenario which contains in a relatively manageable calculation all possible
exotic expectations of the theory. It is immediate to
see, by looking into the mesonic sector, that such  structure
produces spontaneous chiral symmetry breaking. We have
analyzed the nucleon and delta correlators and discovered that
this same mechanism should be reponsible for their mass
difference. We have seen the appearence of exotic
components, both in the nucleon and delta channels, reflecting
the violation (hopefully small) of some crucial symmetries of the
strong interactions, in particular P and CP, without the necessity of
$\Theta$ angle. Since we have not been able to solve $QCD$ exactly, but
have simply postulated an ansatz, it is clear that the non-appearence of
these symmetry violations could not be, in principle, attributable to $QCD$,
but to the inconvenience of our ansatz. In particular a naive quark propagator
structure with only scalar and vector components does not lead to any
symmetry violations.

Finally if confinement would behave in a non-naive manner, as in the
model presented here, our calculation shows that violation of symmetries
may not occur in all channels.
For example, in our case, violation of P only occurs if deltas are
present. Thus one should study effects in different channels
if one wants to gain a full understanding of the
fundamental properties of $QCD$. The advent of high duty cycle
machines might allow for searches, which could be carried out in parallel
with the more traditional neutron experiments.

\section*{Acknowledgement}
We have mantained illuminating discussions with J. Bordes, D.
Espriu and A. Ferrando. We would like to thank M. Asorey for
discussing his work with us. One of us (V.V.) is very much
indepted to D. Espriu for intensive discussions regarding
reference \ci{espriu}, which have been crucial to the present
development. V. Vento would like to thank the members of the
Institut f\"{u}r Kernphysik der Universit\"{a}t Mainz for their
hospitality.

\end{document}